%% file: dlrepair_main.tex
\documentclass[conference]{IEEEtran}
\IEEEoverridecommandlockouts
\usepackage{cite}
\usepackage{amsmath,amssymb,amsfonts}
\usepackage{graphicx}
\usepackage{textcomp}
\usepackage{xcolor}
\usepackage{booktabs}
\usepackage{listings}
\usepackage{multirow}
\usepackage{caption}
\usepackage{subcaption}
\usepackage{framed}

\usepackage{soul}
\usepackage[utf8]{inputenc}
\usepackage[T1]{fontenc}
\usepackage{microtype}
\usepackage{rotating}
\usepackage{paralist}
\usepackage{tabularx}
\usepackage{multicol}
\usepackage{pbox}
\usepackage{enumitem}	
\usepackage{colortbl}
\usepackage{pifont}
\usepackage{xspace}
\usepackage{url}
\usepackage{tikz}
\usepackage{fontawesome}
\usepackage{lscape}
\usepackage{color}
\usepackage{anyfontsize}
\usepackage{comment}
\usepackage{soul}
\usepackage{gensymb}
\usepackage{multibib}
\usepackage{tcolorbox}
\usepackage{balance}
\usepackage{footmisc}

\usepackage[ruled,linesnumbered]{algorithm2e}
\def\BibTeX{{\rm B\kern-.05em{\sc i\kern-.025em b}\kern-.08em
    T\kern-.1667em\lower.7ex\hbox{E}\kern-.125emX}}

\usepackage{xspace}

\newcommand{\dc}{\textsc{DeepCrime}\xspace} 
\newcommand{\AutoT}{\textsc{AutoTrainer}\@\xspace}

\newboolean{showcomments}
\setboolean{showcomments}{true}         
\ifthenelse{\boolean{showcomments}}
  {\newcommand{\nb}[2]{
  \fbox{\bfseries\sffamily\scriptsize#1}
     {\sf\small$\blacktriangleright$\textit{\textcolor{red}{#2}}$\blacktriangleleft$}
   }
  }
  {\newcommand{\nb}[2]{}
   
  }

\newcommand\new[1]{{\color{black}#1}}

\newcommand\paolo[1]{\nb{Paolo}{#1}}

\newcommand{\COMMENT}[1]{}

\definecolor{codegreen}{RGB}{0, 160, 0}
\definecolor{codered}{RGB}{160, 0, 0}

\begin{document}

\title{Repairing DNN Architecture: Are We There Yet?}

\makeatletter
\newcommand{\linebreakand}{%
  \end{@IEEEauthorhalign}
  \hfill\mbox{}\par
  \mbox{}\hfill\begin{@IEEEauthorhalign}
}
\makeatother

\author{\IEEEauthorblockN{Jinhan Kim}
\IEEEauthorblockA{\textit{School of Computing}\\
\textit{KAIST}\\
Daejeon, Republic of Korea\\
jinhankim@kaist.ac.kr}
\and
\IEEEauthorblockN{Nargiz Humbatova}
\IEEEauthorblockA{\textit{Software Institute}\\
\textit{Università della Svizzera italiana (USI)}\\
Lugano, Switzerland \\
nargiz.humbatova@usi.ch}
\and
\IEEEauthorblockN{Gunel Jahangirova}
\IEEEauthorblockA{\textit{Department of Informatics}\\
\textit{King's College London}\\
London, UK\\
gunel.jahangirova@kcl.ac.uk}
\linebreakand
\IEEEauthorblockN{Paolo Tonella}
\IEEEauthorblockA{\textit{Software Institute}\\
\textit{Università della Svizzera italiana (USI)}\\
Lugano, Switzerland \\
paolo.tonella@usi.ch}
\and
\IEEEauthorblockN{Shin Yoo}
\IEEEauthorblockA{\textit{School of Computing}\\
\textit{KAIST}\\
Daejeon, Republic of Korea\\
shin.yoo@kaist.ac.kr}
}

\maketitle

\begin{abstract}
As Deep Neural Networks (DNNs) are rapidly being adopted within large software
systems, software developers are increasingly required to design, train, and
deploy such models into the systems they develop. Consequently, testing and
improving the robustness of these models have received a lot of attention lately.
However, relatively little effort has been made to address the difficulties 
developers experience when designing and training such models: if the evaluation
of a model shows poor performance after the initial training, what should the
developer change? We survey and evaluate existing state-of-the-art techniques
that can be used to repair model performance, using a benchmark of both
real-world mistakes developers made while designing DNN models and artificial
faulty models generated by mutating the model code. The empirical evaluation shows that
random baseline is comparable with or sometimes outperforms existing
state-of-the-art techniques. However, for larger and more complicated models, 
all repair techniques fail to find fixes. Our findings call for further research
to develop more sophisticated techniques for Deep Learning repair.

\end{abstract}


\input{introduction}
\input{background}

\input{benchmark}
\input{empirical_study}

\input{results}

\input{discussion}

\input{threats}

\input{related_work}

\input{conclusion}
\input{ack}

\bibliographystyle{IEEEtran}
\balance
\bibliography{IEEEabrv,biblio,newref}

\vspace{12pt}
\end{document}

%% file: introduction.tex

\section{Introduction}
\label{sec:intro}

Deep Neural Networks (DNNs) are rapidly being adopted into large software 
systems due to the significant advances in their performance across multiple 
domains such as image and speech recognition, machine translation, and 
autonomous driving~\cite{Hinton2012aa,Krizhevsky2017aa,Chen2017aa,chen2015deepdriving
,Sutskever:2014aa,Jean2015aa}.
Especially because some of these application domains, such as medical imaging
~\cite{erickson2017machine, varoquaux2022machine} or autonomous driving~\cite{parekh2022review}, are safety-critical, findings 
about failure-inducing inputs~\cite{Riccio2020bg,Guo2018ml,Gambi2019aa} and adversarial examples~\cite{Goodfellow43405,Kura1607} posed serious 
threats, resulting in significant efforts to test~\cite{Tian2018aa, Pei2017qy,
Ma2018aa,Kim2019aa,Huang2017kx} and improve~\cite{Guo2018ml,Gao:2020aa} the 
robustness of DNN models. 

When considering the proposed ways to improve model performance and robustness, most attention was directed 
to the training dataset, looking for effective methods to augment it in order to address
the discovered deficiencies~\cite{Zhang2019zj, Ma2018gf}. 
However, a model may exhibit poor predictive performance (e.g., high prediction errors)
because of issues affecting the model structure and the training process, not the training data.
Relatively little work has been done to repair the model structure or to improve the training process, 
as compared to re-training the model on an augmented dataset. 
In the following, we refer to \emph{model architecture faults} with the broad meaning of
mistakes made by developers when specifying the model and its training process
in the source code. Examples of such faults include the choice of an
inappropriate activation function for a layer of the DNN or too large a value 
for the learning rate of the optimiser. Such mistakes can have critical impacts 
on the model's predictive performance yet also remain easy to make for the software 
engineers who are not necessarily experts on deep learning~\cite{Humbatova2020kt}. More importantly, these 
mistakes are often not easy to fix manually for  developers, e.g., due to the lack 
of expertise, the stochastic nature of DNN models, and the cost of training and 
evaluating candidate patches. 

This paper assesses existing DNN improvement techniques proposed within both the 
software engineering (SE) and the 
machine learning (ML) research communities. In particular, the former aim explicitly at 
detecting and eliminating problematic symptoms observed during  training (such as vanishing 
gradients or dying ReLU~\cite{Cao2022zz,wardat2022deepdiagnosis}). The latter aim at 
optimising the model's hyperparameters.
While neither directly addresses the issue of model architecture faults made by 
developers, they nonetheless can take as input an underperforming model
and can produce output repair actions that fix architectural faults affecting the model.
As the two families of techniques have a large overlap in terms of the model architecture faults that
they address, we consider both in our empirical assessment.

Specifically, we focus our empirical evaluation on  
\AutoT~\cite{autotrainer}, a representative DNN repair tool,
recently presented at the flagship software engineering conference (ICSE), and on 
HEBO~\cite{Cowen-Rivers2022lm} and BOHB~\cite{bohb}, which represent
 state-of-the-art among the Hyperparameter Optimisation (HPO) techniques developed
by the machine learning community. The latter belong to the Bayesian optimisation family, which
has been shown to outperform all alternative approaches (e.g., search-based)~\cite{JinSH19}. 
As a sanity check, we include random search as a baseline in our study. 

We use a collection of both real-world and artificial model architecture faults 
to evaluate these repair techniques. The real-world model architecture faults 
have been manually curated from the fault benchmark made available by Cao et 
al.~\cite{Cao2022zz}, who in turn collected them from GitHub issues and 
StackOverflow questions about DNN model underperformance.
The artificial model architecture faults have been created by applying
source-level mutation operators~\cite{NHGJPT21}, which are designed
based on a taxonomy of real-world faults~\cite{Humbatova2020kt}. We consider a model
architecture fault fixed once the improvement in model performance, measured
across multiple runs, is statistically significant. The
models we study include image classifiers for MNIST~\cite{LeCun2010vg} and
CIFAR10~\cite{Krizhevsky2009kt}, a text classifier for
Reuters~\cite{reutersdataset}, and eye gaze direction predictors based on the UnityEyes simulator~\cite{unityeyes}.


Our results show that while existing techniques are capable of improving models
with architecture faults, there is ample room for improvement. Surprisingly, the
random baseline generally performs competitively against more sophisticated
repair techniques. Also, both Random and HPO techniques significantly outperform
\AutoT. However, none of the studied techniques shows good performance for
larger and more complex models. A further analysis that simulates different time
budgets for each technique reveals that Random and HPO techniques tend to
perform better when larger budgets are allowed, while \AutoT does not benefit
from larger budgets. Lastly, a complexity analysis of generated patches shows
that all techniques tend to produce more complex patches when compared to the
human-generated ground truth ones (i.e., they have redundant changes compared to
the ground truth patches).

The contributions of this paper are as follows:

\begin{itemize}
\item We present a wide empirical evaluation of existing state-of-the-art techniques for 
the automated repair of DNN model architecture faults. 

\item We provide a carefully curated benchmark of repairable model architecture faults for various DNN benchmark datasets and tasks. It includes both real-world faults as well as artificial mutations. 
\end{itemize}

The rest of the paper is organised as follows. Section~\ref{sec:repair} 
formulates the problem of automatically repairing DNN  architecture 
faults and introduces the techniques we evaluate. Section~\ref{sec:benchmark} 
describes the fault benchmark we use for our empirical evaluation. 
Section~\ref{sec:empirical_study} describes the design of the empirical study, the results 
of which are presented in Section~\ref{sec:results}. 
Section~\ref{sec:discussion} discusses the findings obtained from the empirical 
evaluation, followed by threats to validity (Section~\ref{sec:threats}). Section~\ref{sec:related_work} presents the related work, and 
Section~\ref{sec:conclusion} concludes.


%% file: background.tex

\section{Automated DL Repair}
\label{sec:repair}

The taxonomy of real Deep Learning (DL) faults constructed by Humbatova et
al.~\cite{Humbatova2020kt} includes five top-level categories of DL faults: (1) model
faults; (2) GPU usage faults; (3) API usage faults; (4) training faults; and
(5) tensor faults. In this work, we focus on the faults affecting the
\textit{architecture} of the DNN model, i.e., errors made by developers when
choosing the architecture of the model, including its structure, properties
and the training hyperparameters. In the above mentioned taxonomy, the model
architecture faults match entirely the top-level category (1) (\textit{model
faults}), and partially the top level category (4) (\textit{training faults}).
More specifically, among the training faults, we consider the following
subcategories as model architecture faults: optimiser faults, loss function
faults, and hyperparameters faults, \new{while from the same category, faults affecting
the training data, data preprocessing or the training process are out of scope,
as none of the existing DL repair tools can be applied to these faults.}

In summary, by  \textit{model architecture faults}, we mean the following
(sub-)categories of faults from the DL fault taxonomy~\cite{Humbatova2020kt}: faults
affecting the structure and properties, faults affecting the DNN
layer properties and activation functions, faults due to
missing/redundant/wrong layers, and faults associated with the choice of
optimiser, loss function and hyperparameters (e.g., learning rate, number of
epochs).
Given a DNN model affected by an architectural fault, we define the DNN
\textit{model architecture repair problem} as the problem of finding an
alternative configuration of the model architecture that can improve the model
performance (e.g., accuracy or mean squared error) on the test set by a
statistically significant amount.

In the Machine Learning community, the model architecture repair problem was not
addressed directly, but the existing works on Hyperparameter Optimisation (HPO) can be
regarded as approaches to improve an under-performing model, not only to choose
the initial set of hyperparameters. Moreover, the list of hyperparameters being
optimised by HPO techniques is not limited to the learning rate and the number of
epochs: it instead includes layer properties, activation functions, and even the
number of layers/neurons in the DNN structure, effectively covering all
configuration parameters considered in our definition of the DNN model
architecture repair problem. Hence these approaches fall within the scope of our
empirical investigation. In the Software Engineering community, the model
architecture repair problem was  addressed directly by a few recent works, among
which is \AutoT \cite{autotrainer}.

\subsection{Hyperparameter Optimisation (HPO)}


\textit{Hyperparameter tuning or optimisation} is the problem of finding a tuple
of hyperparameter values such that the model trained with such hyperparameters solves the given
task with acceptable performance~\cite{claesenMoor2015hyperparameter}. Often this problem is
solved manually, but the rising popularity of deep learning methods has pushed for
its automation~\cite{claesenMoor2015hyperparameter,
feurerHutter2019hyperparameter, zela2018towards}. 

Along with the manual search, \textit{grid search (full factorial design)} \cite{montgomery2017design} is a simplistic approach to HPO. It is based on setting a finite set of values for each of the hyperparameters and evaluating all the possible combinations in order to find the best-performing one \cite{feurerHutter2019hyperparameter}. This approach is found not be efficient with the growth of  dimensionality of the configuration space, as the number of the required evaluations increases exponentially  \cite{feurerHutter2019hyperparameter} (it should be noted that each evaluation requires full training of the DNN). 

Another simple approach is  random sampling in the hyperparameter space within a provided search budget\cite{bergstra2012random}.  \textit{Random search} was shown to substantially outperform the manual and grid approaches \cite{bergstra2012random}. It proves to be an appropriate baseline for more sophisticated search algorithms as it is, in theory, able to achieve optimal or nearly optimal results when provided enough budget, and it does not require any knowledge of the function being optimised \cite{feurerHutter2019hyperparameter}.



\setlength{\algomargin}{4mm}

\begin{algorithm}[t!]

\DontPrintSemicolon
\linespread{0.78}\selectfont
\footnotesize
\SetAlCapHSkip{0em}
\SetKwProg{Fn}{function}{}{}
\SetKwInOut{Input}{input}
\SetKwInOut{Output}{output}

\Input{
$D$, initial dataset of model configurations, consisting of pairs  $\langle x_i, y_i\rangle$, with $x_i$ a tuple of hyperparameter values and $y_i$ the model accuracy when trained with $x_i$ \\
$N$, total number of allowed model trainings}
\Output{$x^*$, the best hyperparameter tuple found during the search}
\BlankLine
\Fn{\textsc{\textbf{BO-optimisation}}}{

$b \leftarrow | D |$ \;
\While{$b < N$} {
     $f$ $\leftarrow$ fitSurrogate($D$)\;
     $X$ $\leftarrow$ sampleHyperparameters($f$, $D$) \;
     $b \leftarrow b + |X| $ \;
     $D$ $\leftarrow$ $D$ $\cup$ trainEvaluateModel($X$) \;
}
$x^* \leftarrow \arg\max_{x_i} \{ y_i | \langle x_i, y_i\rangle \in D\}$ \;
\Return {$x^*$}
}
\caption{Bayesian optimisation HPO technique} \label{algo:bo}
\end{algorithm}

\textit{Bayesian optimisation (BO)} is an efficient, state-of-the-art strategy for  global optimisation of objective functions that are costly to evaluate \cite{feurerHutter2019hyperparameter, brochu2010tutorial}. The iterative approach behind BO, shown in Algorithm \ref{algo:bo},  is based on two main components: a probabilistic surrogate model (line 4) and an acquisition function (line 5) \cite{feurerHutter2019hyperparameter}. The surrogate model $f$ (usually a Gaussian process) approximates the objective function (i.e., the model accuracy given the hyperparameters to be used for training) from  the  historical observations made during the previous iterations \cite{feurerHutter2019hyperparameter, Cowen-Rivers2022lm}. 
This function is trained on the available dataset $D$ of previously observed pairs $\langle x_i, y_i\rangle$, where $x_i$ is a tuple of hyperparameter values and $y_i$ is the model accuracy when trained with $x_i$.
Function $f$ is also supposed to provide an estimation of the \textit{uncertainty} affecting its prediction, which is also used to guide the exploration of the configuration space \cite{Cowen-Rivers2022lm}. The acquisition function, \textit{sampleHyperparameters} at line 5, in its turn, calculates the utility of various new candidate tuples by using the surrogate model's prediction (i.e., its estimation of the objective function value) and uncertainty, trading off exploration vs exploitation in the search space\cite{feurerHutter2019hyperparameter, Cowen-Rivers2022lm, brochu2010tutorial}. 
The result is a set of new tuples $X$ that are predicted to bring high model accuracy (exploitation) or diversify the search w.r.t. the previously considered  configurations $D$ (exploration).
When the budget $N$ of allowed model trainings is over, the algorithm returns the configuration $x^*$ associated with the best-performing model obtained after training  with hyperparameters $x^*$.
BO techniques are efficient w.r.t. the number of model trainings and evaluations they require \cite{jones2001taxonomy, sasena2002flexibility}, and produced prominent results in the optimisation of DL network hyperparameters in different  domains\cite{snoek2012practical, snoek2015scalable, melis2017state, dahl2013improving}.

\textit{HEBO (Heteroscedastic Evolutionary Bayesian Optimisation)} is a state-of-the-art BO algorithm developed specifically to optimise a performance metric (validation loss) over the configuration space of various hyperparameters of DL algorithms \cite{Cowen-Rivers2022lm}. The approach won the NeurIPS 2020 annual competition that evaluates black-box optimisation algorithms on real-world score functions \cite{Cowen-Rivers2022lm}. The motivation behind HEBO is the observation that the majority of BO implementations adopt only one acquisition function and Gaussian noise likelihood as a surrogate model to predict the objective function values for candidate hyperparameter tuples \cite{Cowen-Rivers2022lm}. By analysing the available competition data, the authors found out that different acquisition functions provide conflicting results, and noise processes are heteroscedastic and complex. 
To mitigate these issues, HEBO handles heteroscedasticity and non-stationarity of the complex noise processes through non-linear input and output transformations. Moreover, it uses multi-objective acquisition functions with evolutionary optimisers that avoid conflicts by reaching a consensus among different acquisition functions \cite{Cowen-Rivers2022lm}.
Another popular family of HPO approaches, called \textit{bandit-based strategies} \cite{feurerHutter2019hyperparameter,jamieson2016non,hyperband}, has been recently combined with BO, achieving promising results. The main representative of these combined approaches is BOHB \cite{bohb}.

For our evaluation, we chose \textit{Random Search} as the baseline approach and compared it with the two best-performing state-of-the-art HPO algorithms, HEBO and BOHB.

\subsection{AutoTrainer}

\AutoT \cite{autotrainer} is an approach that aims to detect and repair potential DL training problems. It takes as an input a trained DL model saved in the ``.h5'' format and a file that contains training configurations of the model such as optimisation and loss functions, batch size, learning rate, and training dataset name.  Given a DL model and its configuration, \AutoT starts the training process and records training indicators, such as accuracy, loss values, calculated gradients for each of the neurons. It then analyses the collected values according to a set of pre-defined rules and recognises potential training problems. In its current version, the supported symptoms of training problems are: vanishing and exploding gradients, dying ReLU, oscillating loss and slow convergence.  

Once a problem has been detected, \AutoT applies its own built-in repair solutions one by one based on a default order, if an alternative, preferred order is not specified, and checks whether the problem has been fixed by the built-in solution.  The list of predefined solutions includes adding batch normalisation layers,  adding gradient clipping, adjusting batch size and learning rate, substituting activation functions, initialisers and optimisation functions. It should be noted that when applying the possible repair solutions, \AutoT does not re-train the model with the applied repair from scratch, but starts from the already trained initial model and continues the training process for more epochs with the applied solution. If none of the solutions can fix the problem, \AutoT reports its failure to find a repair to the user.

%% file: benchmark.tex
\section{Benchmark}
\label{sec:benchmark}

To evaluate techniques applicable to the problem of DL repair, we prepare a set
of faulty models. In the set, we include programs of two different kinds: those
with artificially seeded faults and faulty programs affected by real faults. In
this section, we describe the nature of such programs, the differences between
the two categories, and the methodology behind the construction of the
benchmark.

\subsection{Artificial Faults}

\begin{table*}[htb]
    \caption{Benchmark of artificial faults}
    
    \begin{small}
    \begin{center}
    \scalebox{0.7}{
    \begin{tabular}{clllcccccc} \hline
    \multirow{2}{*}{Op} & \multirow{2}{*}{Description} & Coverage & Coverage & \multirow{2}{*}{MN} & \multirow{2}{*}{UE} & \multirow{2}{*}{CF10} & \multirow{2}{*}{AU} & \multirow{2}{*}{UD} & \multirow{2}{*}{RT} \\ 
    
    & & HPO-9 & \AutoT & & & & & & \\
    
    \hline
    HLR & Decrease learning rate & Y & Y & \checkmark & \checkmark & - & - &- & \checkmark \\
    HNE & Change number of epochs & Y & N & - & \checkmark & \checkmark & \checkmark &- & - \\
    ACH & Change activation function & Y & Y & - & \checkmark & \checkmark & - &- & \checkmark \\
    ARM & Remove activation function & Y & N & \checkmark & - & - & - &- & \checkmark \\
    AAL & Add activation function to layer  & Y & Y & - & \checkmark & - & - &- & -\\
    RAW & Add weights regularisation & N & N & - & \checkmark & - & - &- & \checkmark \\
    WCI & Change weights initialisation & Y & Y & \checkmark & \checkmark & \checkmark & - &- & \checkmark \\
    LCH & Change loss function & Y & N & - & \checkmark & - & \checkmark & \checkmark & \checkmark \\
    OCH & Change optimisation function & Y & Y & - & \checkmark & - & - & \checkmark & \checkmark \\
    \hline
    \end{tabular}
    }
    \end{center}
    \end{small}
    
    \label{tab:AFmodels}
    \end{table*}

Artificial faults, also known as mutations, are at the core of the software
testing approach called Mutation Testing (MT)~\cite{Jia:2011nx}: a test suite
is deemed mutation-adequate if it can expose the artificially injected faults
(i.e., it can \textit{kill the mutants}). 
As DL systems significantly differ from traditional software, syntactic
mutations are ineffective for mutation testing of DL
\cite{jahangirovantonella}. Thus, over recent years researchers have
proposed a variety of DL-specific mutation operators. Two main groups are
distinguished: pre-training mutation operators and post-training operators. The
post-training operators are applied to a model after the training process is
successfully finished. They focus on altering the structure or weights of the
trained model~\cite{munn, deepmut, deepmut++}. An example of such operators
could be deleting a random layer or adding gaussian noise to a randomly selected
subset of the weights. However, such operators are not realistic
and were found not to be sufficiently sensitive to changes in the test set
quality~\cite{NHGJPT21}. On the other hand, such mutations are fast to
generate and can be preferable in settings with limited time and resources.
Another group, the pre-training operators, seed faults into a model before the
training process begins~\cite{deepmut, NHGJPT21}. They can affect different
aspects of a DL model, such as training data, model architecture and various
hyperparameters. Such operators were shown to be more sensitive to the quality
of test data than those of the post-training mutations~\cite{NHGJPT21}.
A recent DL mutation tool, \dc, generates a set of pre-training
mutants given an original DL system as input~\cite{NHGJPT21}. It is
based on existing, systematic analyses of real faults affecting DL
models~\cite{Humbatova2020kt, islam, zhang}. 
We decided to adopt \dc to generate mutated models for the purposes of our evaluation, as
it produces mutants that are inspired by faults reported by developers to occur
in real life.

The replication package of \dc~\cite{deepcrimeReplication} comes with a
set of pre-trained and saved mutants that cover a range of diverse DL tasks.
Specifically, \dc was applied to a model for handwritten digit
classification  based on the MNIST dataset~\cite{mnist} (MN), to a predictor of
the eye gaze direction from an eye region image~\cite{unityeyes} (UE or UnityEyes), to a
self-driving car designed for the Udacity challenge (UD), to a model that recognises the speaker from an
audio recording (AU), to an image classifier for the CIFAR10 dataset~\cite{cifar10} (CF10), and to a Reuters news categorisation model~\cite{reutersdataset} (RT).

In total, the faulty model dataset of \dc consists of 850 distinct mutants. We examined all of them and selected the mutants that were killed by the test dataset provided with the subjects, according to the statistical mutation killing criterion proposed by Jahangirova and Tonella \cite{jahangirovantonella}, which requires a statistically significant drop in prediction accuracy when the mutant is used to make predictions on the test set. In our evaluation, we adopt this statistical notion of fault exposure, with the parameters suggested by \dc's authors \cite{NHGJPT21}: $p$-value $< 0.05$ and non-negligible effect size.
First of all, out of the pool of the selected mutants, we have excluded those that were generated with the help of mutation operators that affect training data, such as, for example, removing a portion of the training data or adding noise to the data, as these are not model architecture faults. After evaluating the remaining mutants, we introduced thresholds on the performance drop to filter out the mutants that are potentially too easy to detect and repair (have a dramatic drop in performance metric when compared to the original) or those that could be too hard to repair (have a performance comparable to the original one, despite the statistical significance of the difference).
Specifically, we discarded mutants that have an average accuracy lower than 10\% of the original model's accuracy and those that are less than 15\% worse than the original. As for the regression systems, we kept the mutants that have an average loss value between 1.5 and 5 times of the original model's loss. 

When more than one mutant was left after filtering for a given mutation
operator, we have randomly selected one per dataset for inclusion in the final
benchmark. For example, if for the ``change optimisation function operator'', we
were left with two suitable mutants of the MNIST model, which were obtained by
changing the original optimiser to either SGD or Adam~\cite{kingma2014adam}, we took only one of them
randomly. After applying the described filtering procedure, we were left with 25
faulty models suitable for repair. 
As a result, our benchmark contains 25 artificial DL faults split by nine mutation operators
(Op), as shown in Table~\ref{tab:AFmodels}. We also report whether these
fault types are in the scope of the  DL repair tools considered in the empirical
study (columns 3-4, where HPO-9
is a single column for both HEBO and BOHB, configured with a limited set of nine repair operators), as well as the
datasets affected by these faults (columns 5-10).
However, we note that our empirical evaluation excludes two artificial faults from both AU and UD, respectively, because a
single experiment on them with HPO techniques and Random exceeds 48 hours.
The two `Coverage' columns show that the overall coverage of patched fault types by \AutoT is lower than that achieved by HPO
techniques. In addition, the RAW type of fault is not covered by any considered
technique. Still, we include it in the benchmark because an alternative patch,
which differs from the ground truth but is equally effective, could be, in
principle, found by the the repair tools.

\subsection{Real Faults}


\begin{table*}[htb]
    \caption{Benchmark of real faults}
    \begin{small}
    \begin{center}
    \scalebox{0.7}{
    \begin{tabular}{ccclllll} \hline
    \multirow{2}{*}{Id} & \multirow{2}{*}{SO Post \#} & \multirow{2}{*}{Task} & \multirow{2}{*}{Faults} & Coverage & Coverage & \# Hyper \\ 
    
    & & & & HPO-9 & \AutoT & parameters \\
    
    \hline
    D1 & 31880720 & C & Wrong activation function & Y & Y & 15 \\
    \hline
    
    \multirow{2}{*}{D2} & \multirow{2}{*}{41600519} & \multirow{2}{*}{C} & Wrong optimiser | Wrong batch size & Y | Y & Y | Y & 20  \\
    & & & Wrong number of epochs & Y & N &  \\
    \hline
    
    \multirow{2}{*}{D3} &  \multirow{2}{*}{45442843} & \multirow{2}{*}{C} & Wrong optimiser | Wrong loss function | Wrong batch size  & Y | Y | Y & Y | N | Y & 13  \\
    & & & Wrong activation function | Wrong number of epochs  &  Y | Y &  Y | N  &   \\
    
    \hline
    D4 & 48385830 & C & Wrong activation function | Wrong loss function | Wrong learning rate & Y | Y | Y & Y | N | Y & 12  \\
    
    \hline
    D5 & 48594888 & C & Wrong number of epochs | Wrong batch size & Y | Y & N | Y & 18  \\
    \hline
    
    \multirow{2}{*}{D6} & \multirow{2}{*}{50306988} & \multirow{2}{*}{C} & Wrong learning rate | Wrong number of epochs & Y | Y & Y | N & 12  \\
    & & & Wrong loss function | Wrong activation function & Y | Y & N | Y  &   \\
    \hline 
    D7 & 51181393 & R & Wrong learning rate & Y & Y & 9 \\
    \hline
    D8 & 56380303 & C & Wrong optimiser | Wrong learning rate & Y | Y & Y | Y & 17  \\
    \hline
    D9 & 59325381 & C & Wrong preprocessing | Wrong activation function | Wrong batch size & N | Y | Y & N | Y | Y & 19  \\

    \hline
    \end{tabular}}
    \end{center}
    \end{small}
    \label{tab:RFmodels}
    \end{table*}

To enhance our dataset of artificial faults with real-faulty models, we analyse
the benchmark of \textit{DeepFD}, an automated DL fault diagnosis and localisation
tool~\cite{Cao2022zz}. Their benchmark contains 58 buggy DL models collected from
StackOverflow (SO) and GitHub, and provides an original and repaired version
of the DL programs. We first checked if the reported faulty model, its training
dataset, the fault, and its fix correspond to the original SO post or GitHub
commit. We then tried to reproduce  such faults and discarded the issues where
it was not possible to expose the fault in the buggy version of the model or get
it eliminated in the fixed version (i.e., there is no statistically significant
performance difference). 
As a result of such a filtering procedure, we were left with nine real faults,
all coming from SO. \new{Despite our best efforts, we could not collect more faults due to the
rigorousness of the filtering procedure we applied.}
The list of these faults, along with the SO post ID,
fault description, coverage, and the number of hyperparameters by the DL repair
tools considered in our empirical study, is available in Table
\ref{tab:RFmodels}. We can notice that overall the coverage of patched fault
types by \AutoT is lower than that achieved by HPO techniques. Of these nine
models, eight are aimed at solving a classification task (`C' in column 2), and
one is for a regression problem (`R' in column 2). 






%% file: empirical_study.tex

\section{Empirical study}
\label{sec:empirical_study}

\subsection{Research Questions}
The \textit{goal} of our empirical study is to compare existing DL repair tools on our benchmark of artificial and real faults.
We design the empirical study to investigate the following four research questions:

\begin{itemize}
    \item \textbf{RQ1. Effectiveness}: Can existing DL repair tools generate
    patches that improve the evaluation metric? Which repair tool produces the
    best patches?
    \item \textbf{RQ2. Stability}: Are the patches generated by existing DL
    repair tools stable across several runs?
    \item \textbf{RQ3. Costs}: How much does the performance of the repair tools
    change when having a smaller or bigger budget?
    \item \textbf{RQ4. Patch Complexity}: How complex are the
    generated patches? Do they match the ground truths?
\end{itemize}

\subsection{Selected Repair Operators}
\label{sec:top9_ops}
While the number of possible repair combinations grows exponentially with the number of hyperparameters that can be changed, not all repair operators are equally likely to be effective and useful in practice. 
To identify which hyperparameters should be given high priority while searching for a DL repair operator, 
we analyse the taxonomy of real faults in DL systems~\cite{Humbatova2020kt}. 
Specifically, we consider the number of issues coming from SO, GitHub and interviews that
contributed to each leaf of the taxonomy and grouped similar fault types together. Given the resulting list of fault types
sorted by prevalence, we only consider the top ten entries for the purposes of
this study. However, we have to exclude faults types that would typically
lead to a crash, as they are out of scope when considering model architecture faults.
For example, we exclude fault types related to wrong input or output shapes of a layer. This leaves us with the nine most frequent faults. The selected fault categories include: change loss function,
add/delete/change a layer, enable batching/change batch size, change the number
of neurons in a layer, change learning rate, change number of epochs,
change/add/remove activation function, change weights initialisation, and change
optimisation function.


\subsection{Implementations \& Experimental Settings}

We use the Ray Tune~\cite{liaw2018tune} library to implement the Random
baseline, as well as HEBO and BOHB. We set the nine chosen repair operators as the
hyperparameter search space, and we change the time budget to simulate different
experimental settings. Except for Random, the two HPO techniques start the
search from the initial configuration of the faulty model. 
We use a publicly available version of \AutoT.\footnote{\url{https://github.com/shiningrain/AUTOTRAINER}} Our goal is to apply
\AutoT to all of our subject systems. However, its current implementation does
not support regression systems. As a result, \AutoT is applicable to 13
mutants out of 21 and eight real faults out of nine. 

As the performance of DL repair tools can be highly affected by the time budgets, we
run all experiments on three different time budgets, 10, 20, and 50, which are the multipliers of the
training time of initial faulty model.
We run each tool ten times to handle the randomness of the search and the
training process, and report the average of the results. In addition, we split
the test set into two parts: one for guiding the search (i.e., only used during
the search to evaluate candidate patches) and the other for the final evaluation
of the generated patches at the end of the search. Note that \AutoT operates 
differently from HPO techniques: it only begins the repair once it diagnoses a 
failure symptom and continues until it does not observe any. This makes 
it challenging to apply the same time budget configurations as for the other HPO 
techniques. Instead, we simply 
execute \AutoT repeatedly until the total execution time reaches the maximum 
budget, and collect results for lower budgets by looking at the executions 
completed within the time budget. Consequently, a single \textit{run} of \AutoT for 
a given time budget may actually include multiple runs of the tool.


\subsection{Statistical Test \& Evaluation Metrics}

To measure the statistical significance of the patches in terms of their performance metric
values, we use a non-parametric Wilcoxon-signed rank test. The null hypothesis is
that the medians of two lists of metrics values (one from the faulty model and the
other from the patch) are the same, and the alternative hypothesis is that the
medians are different. We use a significance level of 0.05 to reject the null
hypothesis.

Furthermore, we use the following metric, named Improvement Rate (IR), to measure how much the
evaluation metric of the fault ($M_{fault}$) has been improved by the patch, in comparison with the ground truth improvement:
\begin{equation}
    IR = \frac{M_{patch} - M_{fault}}{M_{fix} - M_{fault}}
\end{equation}
where $M_{patch}$ is the evaluation metric of the patch generated by the repair
tools and $M_{fix}$ is the evaluation metric of the ground truth fixed model,
either provided by developers (real faults) or  computed on the model before
mutation (artificial faults). For example, if IR is 1, the generated patch
is as effective as the ground truth fix (it can be noticed that, in principle, IR
can be even greater than 1). We reverse the sign of the IR when
computing it for mean squared error or loss (as lower loss is better).

To quantify the stability of each DL repair tool, we measure the standard
deviation $\sigma$ of the optimal model performance achieved in ten runs of the
tools. 

The complexity of a patch is computed as the number of different
hyperparameters between the generated patch and the initial faulty model. For 
example, if the patch only changes the batch size from 8 to 32, while all 
remaining hyperparameters are unchanged, the patch is considered to have a 
complexity of 1. We normalise the complexity metric by dividing it with the 
total number of hyperparameters, so that it ranges between 0 (i.e., it has the 
same hyperparameters as the initial faulty model)
and 1 (i.e., all hyperparameters have been changed).

Lastly, to quantify the similarity between the sets of repair operators used by the
generated patch and the ground truth, we adopt the Asymmetric Jaccard (AJ) metric for
the repair operators, which measures the number of ground truth repair
operators ($OP_{fix}$) that also appears in the patch ($OP_{patch}$):
\begin{equation}
    AJ = \frac{| OP_{patch} \cap OP_{fix} |}{| OP_{fix} |}
\end{equation}

%% file: results.tex

\section{Results}
\label{sec:results}

\begin{table}[ht]
  \centering
  \caption{Evaluation metric (average: $\mu$; standard deviation: $\sigma$) of
  faulty model, models patched by Random, \AutoT (AT), HEBO, BOHB, and ground
  truth value (Fix) } 
  \label{tab:main_results}
   \scalebox{0.8}{
  \begin{tabular}{l|r|rr|rr|rr|rr|r}
    \toprule
    Id & Faulty & \multicolumn{2}{c|}{Random} & \multicolumn{2}{c|}{AT} & \multicolumn{2}{c|}{HEBO} & \multicolumn{2}{c|}{BOHB} & Fix \\
     & Model & $\mu$ & $\sigma$ & $\mu$ & $\sigma$ &  $\mu$ & $\sigma$ &  $\mu$ & $\sigma$ & \\
    \midrule    
    D1 & 0.52  & \textbf{1.00} & 0.00  &  T/O &  T/O & \textbf{0.76} & 0.24  & \textbf{0.95} & 0.14  & \textbf{1.00} \\
    D2 & 0.53  & \textbf{0.67} & 0.00  & \textbf{0.68} & 0.00 & \textbf{0.67} & 0.00  & \textbf{0.67} & 0.01  & \textbf{0.71} \\
    D3 & 0.61  & \textbf{1.00} & 0.01  & \textbf{0.93} & 0.00 & \textbf{1.00} & 0.00  & \textbf{1.00} & 0.00  & \textbf{1.00} \\
    D4 & 0.10  & \textbf{0.95} & 0.02  & 0.10 & 0.00 & \textbf{0.94} & 0.03  & \textbf{0.93} & 0.06  & \textbf{0.94} \\
    D5 & 0.66  & 0.66 & 0.00  &  N/A &  N/A & 0.66 & 0.00  & 0.66 & 0.00  & \textbf{0.75} \\
    D6 & 0.45  & 0.60 & 0.20  &  T/O &  T/O & \textbf{0.85} & 0.21  & \textbf{0.65} & 0.23  & \textbf{1.00} \\
    \underline{D7} & 6.71  & \textbf{0.91} & 1.73  &  N/A &  N/A & \textbf{2.48} & 2.85  & \textbf{0.49} & 1.02  & \textbf{0.13} \\
    D8 & 0.22  & \textbf{0.57} & 0.03  & \textbf{0.54} & 0.00 & \textbf{0.57} & 0.02  & \textbf{0.57} & 0.02  & \textbf{0.33} \\
    D9 & 0.10  & \textbf{0.13} & 0.03  & 0.10 & 0.00 & \textbf{0.13} & 0.03  & \textbf{0.12} & 0.01  & \textbf{0.99} \\
    \midrule
    C1 & 0.61 & 0.61 & 0.00  & \textbf{0.72} & 0.01 & 0.61 & 0.00  & 0.61 & 0.00  & \textbf{0.70}  \\
    C2 & 0.52 & 0.52 & 0.00  &  N/A &  N/A & 0.52 & 0.00  & 0.52 & 0.00  & \textbf{0.70}  \\
    C3 & 0.49 & 0.49 & 0.00  &  N/A &  N/A & 0.49 & 0.00  & 0.49 & 0.00  & \textbf{0.70}  \\
    \underline{U1} & 0.184 & \textbf{0.152} & 0.051  &  N/A &  N/A & \textbf{0.184} & 0.000  & \textbf{0.184} & 0.000  & \textbf{0.044}  \\
    \underline{U2} & 0.118 & \textbf{0.050} & 0.056  &  N/A &  N/A & \textbf{0.004} & 0.000  & 0.061 & 0.057  & \textbf{0.044}  \\
    \underline{U3} & 0.121 & 0.057 & 0.055  &  N/A &  N/A & \textbf{0.028} & 0.047  & 0.084 & 0.052  & \textbf{0.044}  \\
    \underline{U4} & 0.400 & 0.087 & 0.126  &  N/A &  N/A & \textbf{0.004} & 0.000  & \textbf{0.004} & 0.000  & \textbf{0.044}  \\
    \underline{U5} & 0.071 & 0.071 & 0.000  &  N/A &  N/A & 0.071 & 0.000  & 0.071 & 0.000  & \textbf{0.044}  \\
    \underline{U6} & 0.130 & 0.080 & 0.061  &  N/A &  N/A & 0.080 & 0.061  & \textbf{0.042} & 0.058  & \textbf{0.044}  \\
    \underline{U7} & 0.098 & \textbf{0.023} & 0.037  &  N/A &  N/A & \textbf{0.033} & 0.043  & \textbf{0.033} & 0.043  & \textbf{0.044}  \\
    \underline{U8} & 0.163 & 0.163 & 0.000  &  N/A &  N/A & 0.163 & 0.000  & 0.163 & 0.000  & \textbf{0.044}  \\
    M1 & 0.85 & 0.86 & 0.04  &  N/A &  N/A & \textbf{0.94} & 0.04  & 0.87 & 0.04  & \textbf{0.99}  \\
    M2 & 0.11 & \textbf{0.43} & 0.36  & \textbf{0.99} & 0.00 & \textbf{0.93} & 0.04  & 0.21 & 0.26  & \textbf{0.99}  \\
    M3 & 0.10 & \textbf{0.52} & 0.41  & \textbf{0.31} & 0.05 & \textbf{0.87} & 0.25  & \textbf{0.45} & 0.35  & \textbf{0.99}  \\
    R1 & 0.51 & 0.56 & 0.10  & \textbf{0.58} & 0.00 & 0.52 & 0.03  & 0.52 & 0.02  & \textbf{0.82}  \\
    R2 & 0.29 & \textbf{0.67} & 0.09  & 0.23 & 0.00 & \textbf{0.49} & 0.13  & \textbf{0.71} & 0.10  & \textbf{0.82}  \\
    R3 & 0.35 & \textbf{0.68} & 0.10  & 0.34 & 0.00 & 0.53 & 0.19  & \textbf{0.64} & 0.14  & \textbf{0.82}  \\
    R4 & 0.66 & 0.70 & 0.06  & \textbf{0.81} & 0.00 & 0.67 & 0.03  & \textbf{0.72} & 0.07  & \textbf{0.82}  \\
    R5 & 0.64 & \textbf{0.72} & 0.05  & \textbf{0.82} & 0.00 & \textbf{0.69} & 0.07  & \textbf{0.71} & 0.05  & \textbf{0.82}  \\
    R6 & 0.50 & \textbf{0.75} & 0.04  & \textbf{0.56} & 0.00 & 0.61 & 0.12  & \textbf{0.72} & 0.08  & \textbf{0.82}  \\
    R7 & 0.30 & \textbf{0.68} & 0.13  & 0.12 & 0.00 & \textbf{0.61} & 0.12  & \textbf{0.67} & 0.09  & \textbf{0.82}  \\
    \bottomrule
  \end{tabular}
  }
\end{table}

\subsection{Effectiveness (RQ1)}
\label{sec:RQ1}

Table~\ref{tab:main_results} shows the evaluation metric value (accuracy or
regression loss, depending on the model; regression models are underlined) of
the patched models averaged over ten runs of patch generation ($\mu$) for
Random, \AutoT (AT), HEBO and BOHB. Column `Faulty Model' shows the metric value
for the initial faulty model, while column `Fix' shows the value for the ground
truth repaired model. The cases that show statistical significance of the
difference between the metric value of the faulty model and patched model are
highlighted in \textbf{bold}. The fault Id (first column) is composed of a
letter and an incremented integer. The letter identifies the dataset: D = real
faults, C = CIFAR10, U = UnityEyes, M = MNIST, R = Reuters. `N/A' means that
\AutoT cannot be applied to the faulty program (e.g., to UnityEyes, which is a regression model) or did not find any failure
symptoms, and `T/O' means that \AutoT did not have enough time to find any
patch. Note that,
due to space limits, Table~\ref{tab:main_results} only shows the results for the time budget 20, i.e., 20 times longer than the training time used by the initial, faulty model. For the full
tables, please see the online supplementary material at \url{https://github.com/dlfaults/dnn-auto-repair-empirical-assesment}. 

Overall, ground truth patches (column `Fix' in Table~\ref{tab:main_results})
show the highest evaluation metrics, although there are a few cases where the
Random or HPO find better patches than the ground truth: D8, U2, U3, U4, and U7.

\begin{figure}[ht]
  \centering
  \includegraphics[width=\linewidth]{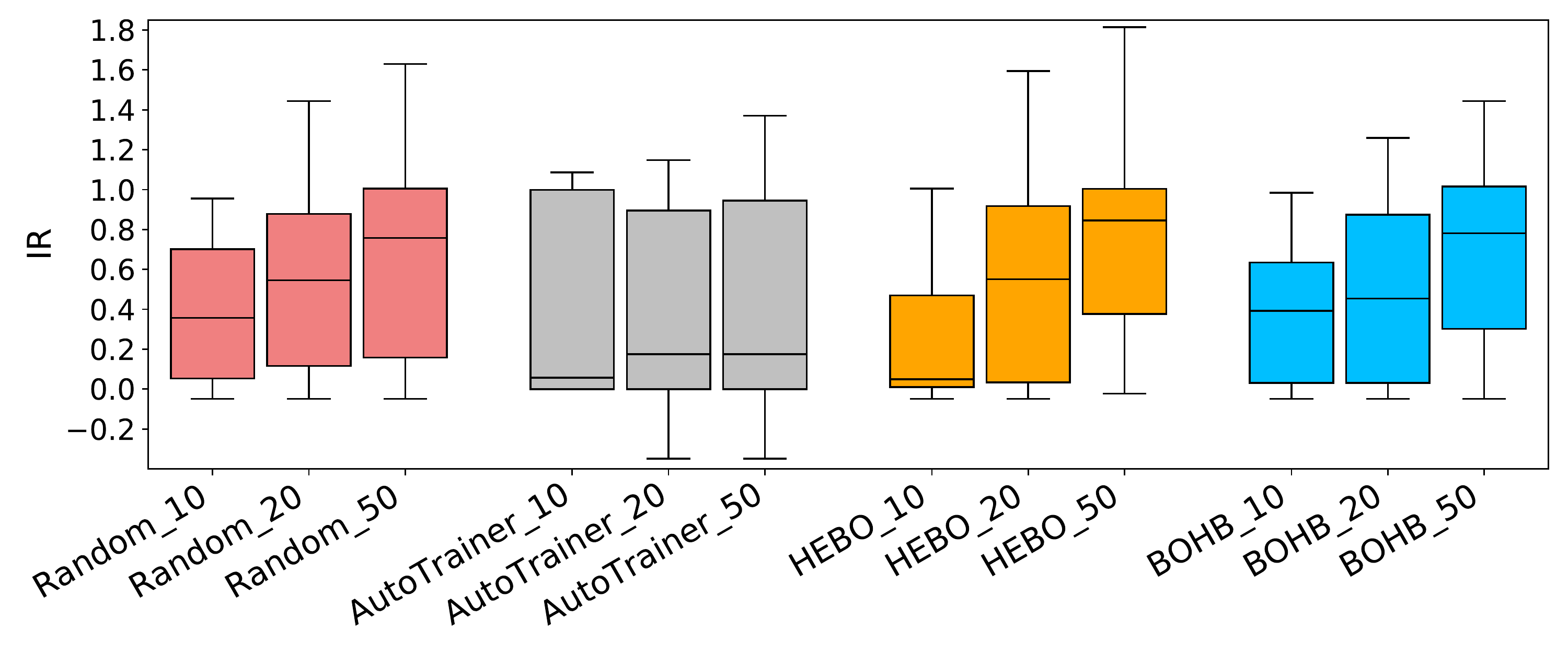}
  \caption{IR values from all faults in the benchmark, broken down by the
   combinations of repair technique and budget, shown as [technique]\_[budget].
   Note that some IR values are higher than 1.0, meaning that the corresponding patches are better than the ground truth patches.}
  \label{fig:IR}
\end{figure}

Next, we compare the repair performance between four repair techniques, based on
the number of statistically significant patches found by each.
Out of the 52 cases,\footnote{For a fair comparison, we only consider the cases where all four techniques can run on the faults without errors. Also, this number of cases represents a comprehensive result by aggregating the results of \textbf{all time budgets.}}
BOHB and HEBO find patches in 35 and 36 cases (67\%, 69\%), respectively, showing statistical significance, followed
by Random with 33 cases (63\%), and \AutoT with 27 cases (52\%). Furthermore,
Figure~\ref{fig:IR} shows IR values for the considered techniques: within the 20
trainings time budget, the median IR values of both Random and HEBO are 0.55, followed by BOHB with
0.45 and \AutoT with 0.18 (see Section~\ref{sec:RQ3} for the
analysis on all budgets). This means that, in general, \AutoT and HPO techniques fail to generate
more effective patches than Random. Despite being a baseline technique, overall, Random
performs surprisingly well in terms of IR, across all subjects and faults.
\new{This conclusion differs from the ones reported in the papers of HEBO and
BOHB~\cite{Cowen-Rivers2022lm, bohb}, which showed that their techniques are
better than Random. We hypothesize that this is due to the different set of subjects that
we considered, which has a larger number of hyperparameters and
correspondingly a larger search space: our study required tuning of an average
of 15 hyperparameters as opposed to the six in their studies.}

\begin{figure}[!ht]
  \begin{subfigure}{\linewidth}
      \centering
      \includegraphics[width=\linewidth]{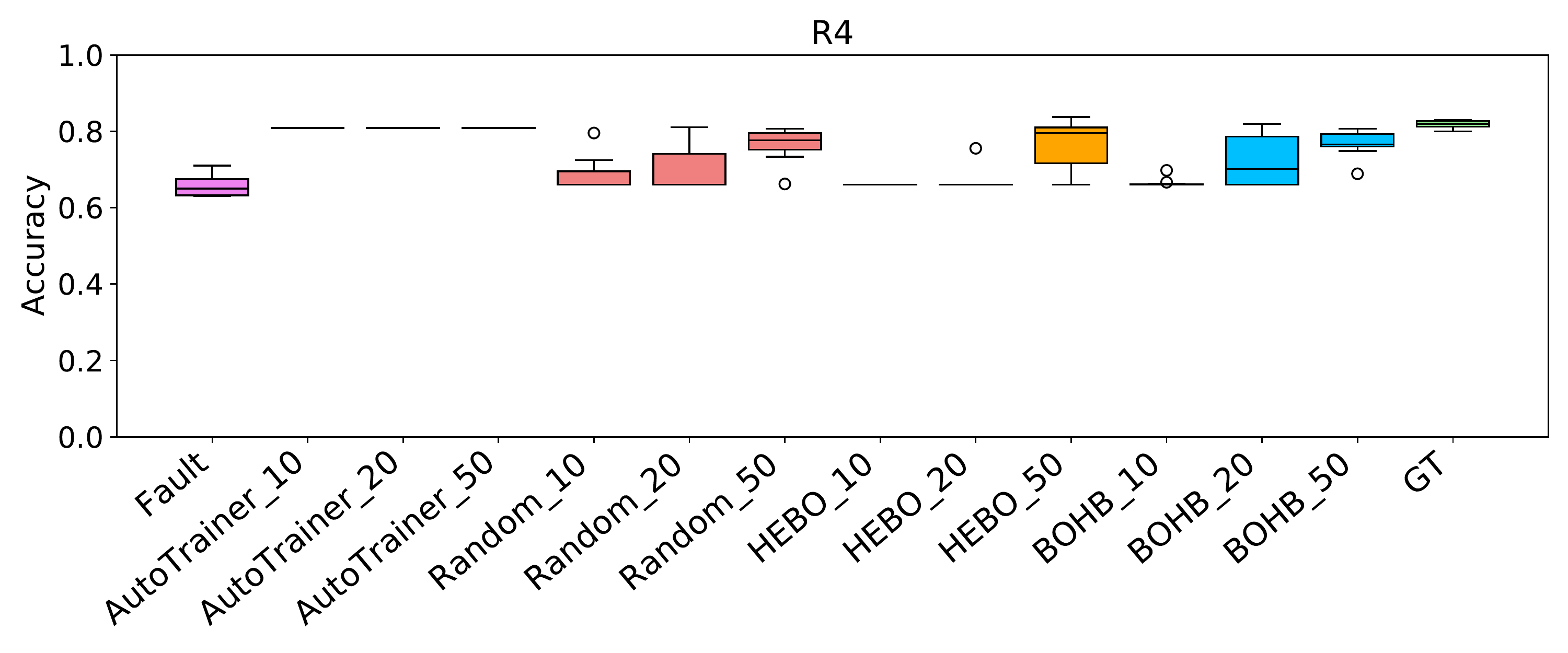}
      \caption{Artificial fault R4}
      \label{fig:example_boxplots1}
  \end{subfigure}
  \begin{subfigure}{\linewidth}
      \centering
      \includegraphics[width=\linewidth]{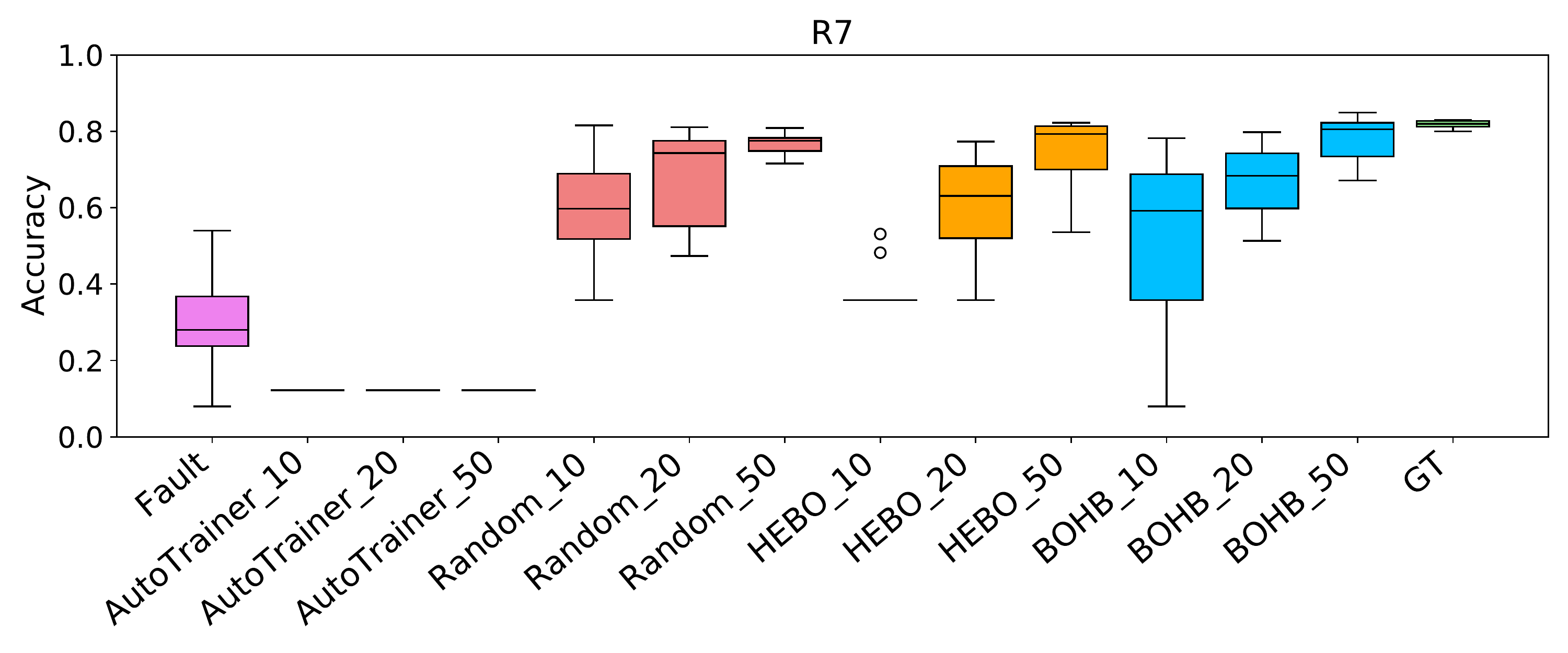}
      \caption{Artificial fault R7}
      \label{fig:example_boxplots2}
  \end{subfigure}
  \caption{ Example boxplots of the results from two artificial faults, showing the
  accuracy of the generated patches. The $x$-axis represents combinations
  of repair techniques and budgets, shown as [technique]\_[budget].}
  \label{fig:example_boxplots}
  \vspace{-0.2cm}
\end{figure}

The boxplot of each fault provides a closer look at how differently each technique performs depending on the type of fault. 
For example, Figure~\ref{fig:example_boxplots} presents the accuracy boxplots for two artificial
faults. As shown in Figure~\ref{fig:example_boxplots1}, \AutoT 
easily and consistently fixes this fault, even with a small budget. 
While \AutoT's coverage of different fault types is not that high, being generally lower than that of HPO techniques (see Tables~\ref{tab:AFmodels} \& \ref{tab:RFmodels}), 
when a fault type is in the scope of \AutoT, it can be effective, especially on simpler
faults, such as mutants generated by \dc, which by construction, can be fixed with a single repair operation. 
Faults M2, R4, and R5 in
Table~\ref{tab:main_results} are  cases in which \AutoT finds good
patches more easily than others. However, as shown in
Figure~\ref{fig:example_boxplots2}, if there is no specific repair operator for
the fault, \AutoT cannot find a good patch. In contrast, since HPO techniques are 
designed to apply multiple repair operators at once, they effectively search 
a wider space of patches and, thus, are more likely to find better patches for
more complicated cases. 
While the overall trend is that the fixes produced by Random are better than
\AutoT and are comparable with fixes by HPO techniques, this is not always the
case. Also, the efficacy of a repair technique depends on the type of faults;
there is no single best repair technique.

\begin{framed}
  \textbf{Answer to RQ1}: In general, random baseline produces comparable or better patches
  than other repair techniques, but the effectiveness of tools varies depending
  on the fault, which justifies the need for future work to find more efficient ways of exploring the hyperparameter space.
\end{framed}

\subsection{Stability (RQ2)}
\label{sec:RQ2}

Table~\ref{tab:main_results} shows the standard deviations ($\sigma$), which quantify
the stability of the patches found by each tool across ten runs (i.e., $\sigma$ quantifies the performance variability
of the best patched model  across multiple executions of each tool).
Below, we comment on the standard deviation of each tool, considering only the cases showing
statistical significance of the model performance improvement.

\AutoT has the smallest average standard deviation of 0.006, followed by HEBO
with 0.060, Random with 0.085, and BOHB with 0.094. \AutoT is shown to be the
most stable technique: this is because, in principle, the number of repair
operators being applied is relatively small compared to the others (see their
coverages in Tables~\ref{tab:AFmodels} \& \ref{tab:RFmodels} and complexities in Section~\ref{sec:RQ4}
for details), allowing it to generate consistent patches across executions,
despite the randomness occurring in multiple runs. In contrast, HPO techniques,
as well as Random, tend to produce more diverse and different patches, which
implies that their patches are less stable in terms of patched model
performance. This calls for future work to improve the stability of the repair
techniques for DL models, especially when the patches are complex in terms of
the number of changed hyperparameters and applied repair operators.

\begin{framed}
  \textbf{Answer to RQ2}: \AutoT produces similar patches across several runs
  since it operates by applying operators selected from a relatively small set, while HPO techniques and
  Random produce varied patches, hence, they are more prone to instability.
\end{framed}

\subsection{Costs (RQ3)}
\label{sec:RQ3}

Automated program repair for traditional software usually requires a significant amount of time and
computational resources, as it needs to search a large space of patches
while running the tests for each candidate patch. Techniques such as Random and HPO also have
similar issues because each patch requires training and validating its model
from scratch.


We investigate three different time limits, 10, 20, and 50, under the
assumption that developers may have different time constraints when repairing a
faulty DL model. Due to lack of space, we report only a time limit of 20 in
Table~\ref{tab:main_results} (full results are available in online supplementary
material). As expected, all techniques produce more patches showing the statistical
significance of the improvements when larger budgets are allowed. For instance, Random finds patches
showing statistical significance in 14 cases with a 10 time budget, which
becomes 17 cases with a 20 time budget and 23 cases with a 50 time
budget. This trend is consistent even considering IR, as shown in
Figure~\ref{fig:IR}: larger time budget results in larger IR as well as a smaller
standard deviation. \AutoT does not take advantage so much of a larger time
budget, compared to the other techniques, due to its limited search space. HEBO
can be a good alternative to Random when the budget is large such as 50: it
shows slightly better performance than Random with a smaller standard deviation.

Overall, given larger budgets, our results support the use of HPO
techniques, such as HEBO, which are preferable to \AutoT because of the
narrower scope of the latter.

\begin{framed}
  \textbf{Answer to RQ3}: For all DL repair techniques, using a larger time budget
  results in more stable and better patches. The results also show that \AutoT does not benefit from larger budgets, while HPO techniques can benefit from them.
\end{framed}

\subsection{Patch Complexity (RQ4)}
\label{sec:RQ4}

\begin{figure}[!ht]
  \begin{subfigure}{\linewidth}
      \centering
      \includegraphics[width=\linewidth]{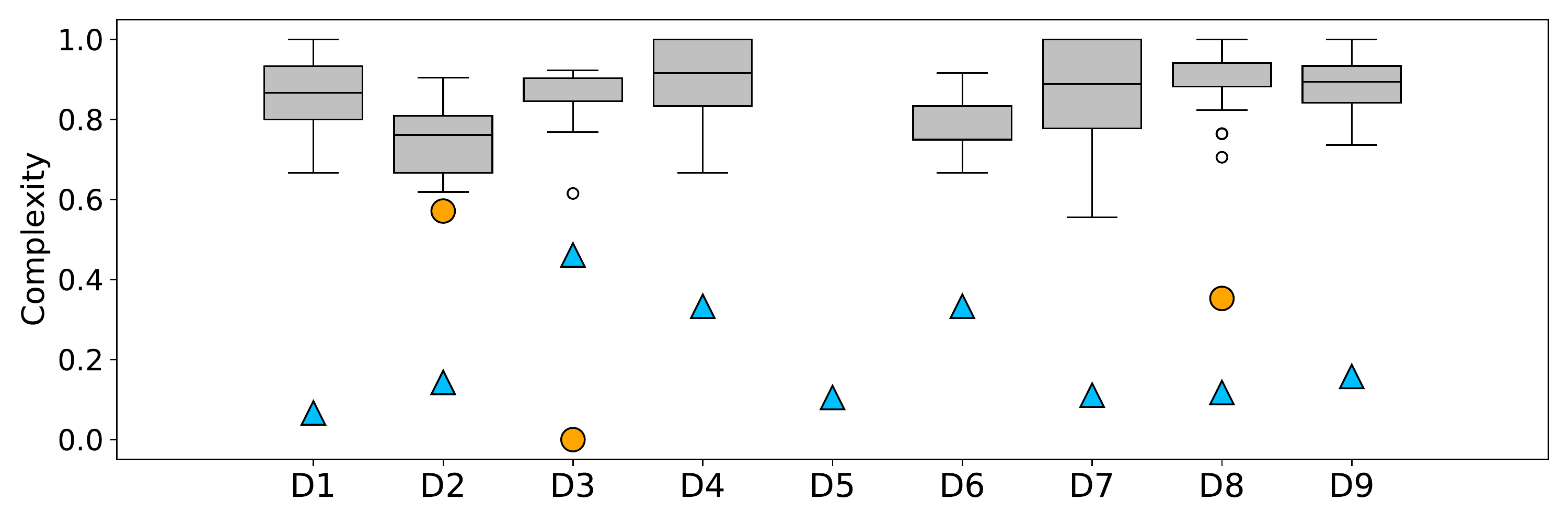}
      \caption{Real faults}
      \label{fig:complexity_plot1}
  \end{subfigure}
  \begin{subfigure}{\linewidth}
      \centering
      \includegraphics[width=\linewidth]{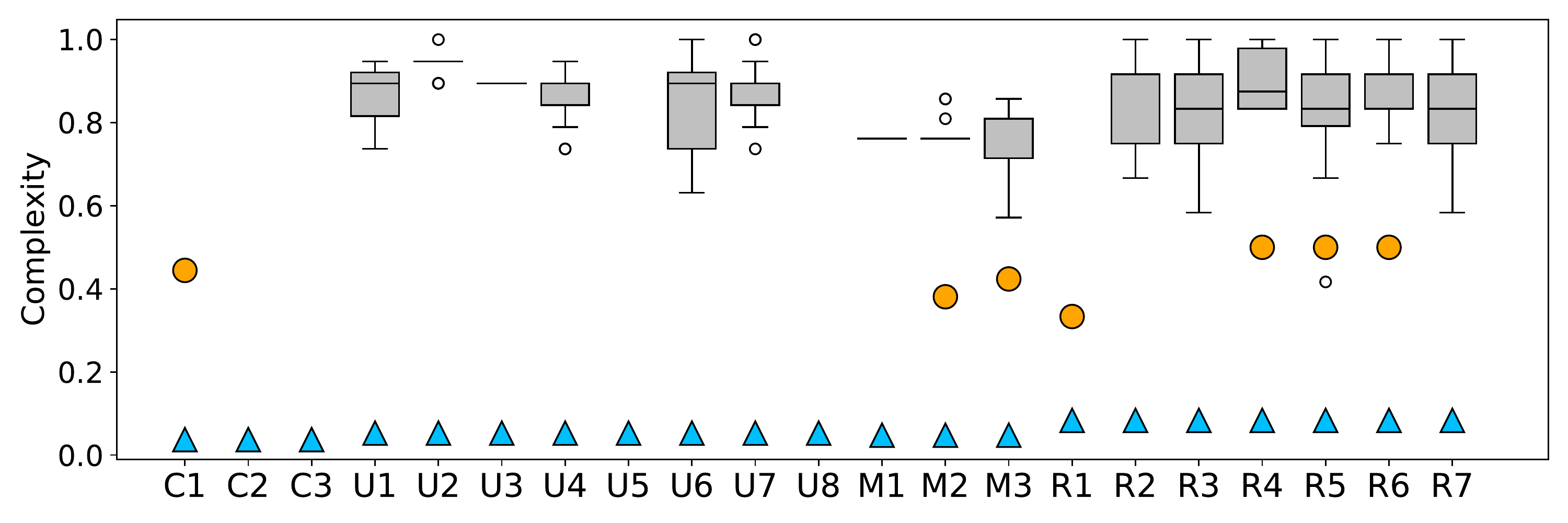}
      \caption{Artificial faults}
      \label{fig:complexity_plot2}
  \end{subfigure}
  \caption{Complexities of statistically significant patches. Boxplots represent HPO and Random's patches\footref{fn:complexity}; triangles the ground truth's one and circles the \AutoT's one.}
  \label{fig:complexity_plot}
\end{figure}
Figure~\ref{fig:complexity_plot} presents the boxplots of the complexity of the
statistically significant patches generated by HPO techniques and Random with 
time
budget 20. The blue triangles and orange circles show the complexity of the
ground truth patches and \AutoT's patches, respectively.\footnote{We present 
integrated results of Random and two HPO techniques as they all show similar trends, and we do not
use boxplots for ground truth and \AutoT as their variance is too small. Also, note that there are missing boxplots and circles because we only consider statistically significant patches.\label{fn:complexity}}
Overall, the complexity of the generated patches of HPO and Random
is much higher than the complexity of the ground truth patches. This means that
the generated patches manipulate many different hyperparameters (around 80\% to
90\% of them) to achieve an improvement of the faulty model. In contrast, a ground
truth patch makes fewer changes, despite achieving similar or higher
evaluation metric values. The main reason for this difference is that both
Random and HPO explore the hyperparameter space at large in search for
configurations that improve the model's accuracy. Random is completely
unconstrained in its exploration: thus, it is expected that it can generate 
solutions that are far from the initial faulty model. HPO, on the other hand, balances exploitation (i.e.,
local improvements of the best model found so far, which at the beginning is the
initial faulty model) and exploration (i.e., it samples new diversified  points
in the hyperparameter space to avoid getting stuck in a local minimum). 
Consequently, the results suggest that, in our subjects, the exploration 
component is dominant, and improvements are obtained only when HPO techniques 
moved away from the initial model. 

In general, the patches generated by \AutoT have lower complexity than Random and HPO. This
is consistent with its design principle: it can handle a narrow set of
repair actions, targeting specific fault types, which makes the tool either
effective and capable of improving the initial solution with a small number
of changes or completely ineffective.

Despite the high complexity, the observed AJ values\footnote{AJ figures are available at \url{https://github.com/dlfaults/dnn-auto-repair-empirical-assesment}} suggest that the generated 
patches do contain the same \emph{ingredients} as the ground truth patches, 
i.e., they include similar repair operators. For real faults, the AJ values for 
HPO and Random have a mean of 0.97 and a standard deviation of 0.11; for 
artificial faults, 0.90 and 0.29, respectively. The AJ values for \AutoT, 
however, reflect its narrower repair scope, with a mean and standard deviation of 
0.26 and 0.19 for real faults, and 0.61 and 0.49 for artificial faults, 
respectively. Considering this, in conjunction with the high complexity values, 
we suggest that the generated patches may be \emph{bloated}, i.e., they contain 
redundant changes when compared to the ground truth. 


\begin{framed}
  \textbf{Answer to RQ4}: The complexity of the patches generated by HPO
  techniques and Random are high compared to the ones of the ground truth
  and of \AutoT, which demands better Bayesian optimisation
  algorithms that can take advantage of the initial, faulty model.  
\end{framed}

%% file: discussion.tex
\section{Discussion}
\label{sec:discussion}


The analysis of the existing benchmarks of real faults currently used in the literature has revealed that the majority of real faults collected so far are rather simplistic. In many cases, the models represent toy examples for naive tasks and data used to train and test the models are either randomly generated or too small. On the other hand, the artificial faults produced by \dc cover a larger variety of fault types and affect more diverse and complex models. 

Indeed, in our empirical evaluation, artificial faults were more challenging to repair than the real ones. The evaluated approaches are either unable to reach the performance of an un-mutated model or show a high standard deviation across repair repetitions. 
A common pattern we observed in the results is that on large models such as
CIFAR10 model, all techniques could not generate any successful fixes (see C2 and C3 in
Table~\ref{tab:main_results}). Since the number of hyperparameters of the
CIFAR10 model is 27, which is twice bigger than that of the Reuters model, the
search space is relatively large, so it becomes more difficult to find patches.
While developing more advanced search techniques to deal with large models
is a promising direction for future work, the other option would be combining
Fault Localisation (FL)~\cite{Cao2022zz, wardat2022deepdiagnosis} and repair
techniques. FL techniques can narrow down the search space  
and pinpoint the locations of a fault (i.e., faulty hyperparameters), which
can be used as a starting point for the repair techniques. See
Section~\ref{sec:related_work_fl} for more details on FL techniques. 

Future work could include various repair operations. We used nine frequent fault types,
but this selection is insufficient to cover all faults in real world. In
particular, the existing repair operations and techniques do not cover faults
related to the quality and pre-processing of training and test data.


Compared to traditional Automated Program Repair (APR) techniques for source
code, one critical step that is missing in model architecture repair is
\textit{patch minimisation}. Although, our analysis shows that smaller patches
do exist and such patches are useful for developers, minimization might be
difficult due to the stochastic nature of model repair. Existing model
slicing~\cite{Zhang2020kz} and pruning~\cite{Liu2020jg} techniques tend to apply
directly to the trained models and not to the source code that defines the model
architecture. Patch minimisation for model architecture faults remains an
unexplored area.


Lastly, our results open up a new direction of research that targets the space
of higher-order patches by smaller and more local changes of the initial, faulty
model. In fact, HPO techniques are designed to start from scratch far from the
initial hyperparameters. Intensification of the search around the initial faulty
model seems a promising research direction.



%% file: threats.tex

\section{Threats to Validity}
\label{sec:threats}

One of the threats to \textbf{internal validity} is the selection of the HPO algorithms.
We carefully studied state of the art in HPO algorithms and chose novel and best-performing
 approaches as well as generally accepted baseline. To avoid incorrect
implementations of those algorithms, we used widely used libraries and frameworks.
The main threat to \textbf{external validity} is the construction of the benchmark used
for the comparison of the approaches. To mitigate any risks, we included both
artificial faults that cover a variety of subjects and a dataset of real faults
used in the previous literature. All the faults included in our benchmark were 
obtained through a methodologically sound selection procedure.
Threats to \textbf{construct validity} lie in a correct measurement of the
performance of the repair tools. All evaluation metrics used in the benchmark are
standard and widely used in the ML literature. For what concerns \textbf{conclusion validity}, we measured effect size using 
our custom metric IR and statistical significance using Wilcoxon's test.

%% file: related_work.tex

\section{Related Work}
\label{sec:related_work}
\subsection{Model-level Repair}
To the best of our knowledge, no automatic source-level repair tool currently
exists that aims to fix the performance of a given faulty DL model by applying
patches to the source code that defines the model's architecture and
hyperparameters. As discussed in Section~\ref{sec:repair}, the closest
approaches come from machine learning, in particular, those solving the HPO
problem, and from software engineering, \AutoT, a tool that continues to train
an already trained model using patched hyperparameters. The goal of our
empirical study was to compare these two families of approaches when adapted to
solve the model architecture repair problem. No previous empirical study
attempted to conduct any similar comparison.

On the other hand, post-training, model-level repair of DNN networks, i.e.,
repair through the modification of the weights of an already trained model, is
gaining increasing popularity. \textit{CARE}~\cite{sun2022causality} identifies
and modifies weights of neurons that contribute to detected model misbehaviors
until the defects are eliminated.
\textit{Arachne}~\cite{Sohn2022cr} operates similarly to \textit{CARE} while
ensuring the non-disturbance of the correct behaviour of a model under repair.
\textit{GenMuNN}~\cite{wu2022genmunn} ranks the weights based on the effect on
predictions. Using the computed ranks, it generates mutants and evaluates and
evolves them using a genetic algorithm.
\textit{NeuRecover}~\cite{tokui2022neurecover} keeps track of the training
history to find the weights that have changed significantly over time. Such
weights become a subject for repair if they are not beneficial for the
prediction of the successfully learnt inputs but have become detrimental for the
inputs that were correctly classified in the earlier stages of the training.
Similarly to NeuRecover, \textit{I-Repair}~\cite{henriksen2022repairing} focuses
on modifying localised weights to influence the predictions for a certain set of
misbehaving inputs, whereas minimising the effect on the data that was already
correctly classified. \textit{NNrepair}~\cite{usman2021nn} adopts fault
localisation to pinpoint suspicious weights and treats them by using constraint
solving, resulting in minor modifications of weights.

PRDNN~\cite{sotoudeh2021provable} took a
slightly different path by focusing on the smallest achievable single-layer
repair. If provided with a limited set of problematic inputs and a model, this
algorithm returns a repaired DNN that produces correct output for these and
similar inputs and retains the model's behaviour for other, dissimilar kinds of
data. \textit{Apricot}~\cite{Zhang2019zj}, however, uses a DL model
trained on a reduced subset of inputs and then uses the weights of the reduced model to
adjust the weights of the full model to fix its misbehaviour on the inputs from
the reduced dataset. In our work, we are interested in the approaches that recommend changes to the
model's source code rather than patching the weights of the model.


\subsection{Fault Localisation}
\label{sec:related_work_fl}

Fault localisation in DNNs is a rapidly evolving area of DL
testing~\cite{wardat2021deeplocalize, wardat2022deepdiagnosis, Cao2022zz,
nikanjam2021automatic, schoop2021umlaut}. Most of the proposed approaches focus
on analysing the run-time behaviour during the model training. According to the
collected information and some predefined rules, these approaches decide whether
they can spot any abnormalities and report them~\cite{wardat2021deeplocalize,
wardat2022deepdiagnosis, schoop2021umlaut}.

During the training of a model, both
\textit{DeepDiagnosis}~\cite{wardat2022deepdiagnosis} and
\textit{DeepLocalize}~\cite{wardat2021deeplocalize} insert a callback that
collects various performance indicators such as loss function values, weights,
gradients and activations. Both tools then compare the analysed values with a
list of pre-defined failure symptoms. \textit{UMLAUT}~\cite{schoop2021umlaut}
combines heuristic static checks of the model structure and its parameters with
dynamic monitoring of the training and the model behaviour. It complements the
results of the checks with the analysis of the error messages, providing best
practices and suggestions on how to deal with the faults. 

Unlike previously discussed methods,
\textit{Neuralint}~\cite{nikanjam2021automatic} is a model-based approach that
employs meta-modelling and graph transformations for fault detection. Given a
model under test, it constructs a meta-model consisting of the base skeleton and
some fundamental properties. This model is then checked against a set of 23
rules embodied in graph transformations, each representing a fault or a design
issue.
\textit{DeepFD}~\cite{Cao2022zz} employs mutation testing to construct a
database of mutants and their original models to train a fault type ML
classifier. From the mutants, it extracts a number of runtime features
and use several combinations of them to localise the faults.

Although all of these approaches are potentially useful for the task of automated repair
of DNNs, they just provide suggestions without any detailed instructions
on how to change the faulty model. Hence, they could not be included in our
empirical comparison of DL repair tools.

%% file: conclusion.tex
\section{Conclusion}
\label{sec:conclusion}

 In this work, we evaluate techniques proposed in the ML and SE research
 communities that are applicable to the problem of repair of DNN architecture faults. In
 particular, we compare the state-of-the-art hyperparameter tuning algorithms
 HEBO~\cite{Cowen-Rivers2022lm} and BOHB~\cite{bohb}, which are based on Bayesian
 optimisation, and the recent DNN repair tool called
 \AutoT~\cite{autotrainer}, while using random search as a baseline. To
 allow a thorough assessment, we apply these techniques to a carefully
 collected benchmark of real and artificial faults. The obtained results
 indicate that all of the evaluated techniques are able to enhance the
 performance of  fault models in some cases but are often not as effective as
 the ground truth fixes. Moreover, the generated patches tend to have a higher
 complexity than that of the ground truth. According to our observations, for
 simpler models, more advanced approaches do not happen to outperform random
 search, whilst random search and HPO algorithms clearly surpass \AutoT. For 
 more  complex models, all considered approaches fail to perform well. 
 Thus, there is ample space for improvement in the area of DNN model 
 architecture repair. Furthermore, our findings reveal a number of promising 
 future research directions: a synergy with DNN fault localisation techniques, 
 the need for more sophisticated repair operators and algorithms, and the need 
 for patch minimisation.

%% file: ack.tex

\section*{Acknowledgement}
Jinhan Kim and Shin Yoo have been supported by the Engineering Research Center
Program through the National Research Foundation of Korea (NRF) funded by the
Korean Government (MSIT) (NRF-2018R1A5A1059921), NRF Grant (NRF-2020R1A2C1013629),
Institute for Information \& communications Technology Promotion grant funded by
the Korean government (MSIT) (No.2021-0-01001), and Samsung Electronics (Grant
No. IO201210-07969-01). This work was partially supported by the H2020 project
PRECRIME, funded under the ERC Advanced Grant 2017 Program (ERC Grant Agreement
n. 787703).